# OOML: Structured Approach to Web Development

John Francisco     Victor Sadikov

AT&T Information Technology Unit
200 S Laurel Ave A5-{3C05, 2D20}
Middletown, NJ 07748, USA
1 (732) 420-{8809, 7543}

{shemp, vic}@att.com

## ABSTRACT
In today's world of Web application development, programmers are commonly called upon to use the Hypertext Markup Language (HTML) as a programming language, something for which it was never intended and for which it is woefully inadequate. HTML is a data language, nothing more. It lacks high level programming constructions like procedures, conditions, and loops. Moreover it provides no intrinsic mechanism to insert or associate dynamic application data. Lastly, despite the visibly apparent structure of a web page when viewed in a browser, the responsible HTML code bears little to no discernible corresponding structure, making it very difficult to read, augment, and maintain.

This paper examines the various drawbacks inherent in HTML when used in Web development and examines the various augmenting technologies available in the industry today and their drawbacks. It then proposes an alternative, complete with the necessary constructs, structure, and data associating facilities based upon server-side, Extensible Stylesheet Language Transforms (XSLT). This alternative approach gives rise to an entirely new, higher level, markup language that can be readily used in web development.

## Categories and Subject Descriptors
I.7.2 [**Document and Text Processing**]: Document Preparation – *markup languages, scripting languages, format and notation, hypertext/hypermedia, languages and systems, standards.* D.2.11 [**Software Engineering**]: Software Architectures – *languages, patterns.* H.5.4 [**Information Interfaces and Presentation**]: Hypertext/Hypermedia – *architectures.* D.2.13 [**Software Engineering**]: Reusable Software – *reusable libraries, reuse models.*

## 1. INTRODUCTION
The Hypertext Markup Language (HTML) is a data language designed to present mostly static data inside a generic user interface application or browser.

It was never intended to be a programming language, and significant portion of web based development centers around compensating for this deficiency. Moreover, most web applications have a need to associate dynamically generated data with the web page being displayed, that is, few web pages are truly static. Since HTML does not have an intrinsic mechanism for including dynamic data, an equal amount of web based development centers around the association of dynamic data with a web page. This paper outlines the drawbacks of HTML, surveys the prevailing technologies that exist to compensate for those drawbacks, and then presents an alternate approach to compensate for lacking text processing and data manipulating constructions. The proposed additions to HTML are implemented as a library in Extensible Stylesheet Language.

### 1.1 The Drawbacks of HTML
The Hyper Text Markup Language was never intended as a programming language, this is its major drawback. Moreover, because of its rather limited constructs, HTML code tends to lack structure, even though such structure is readily apparent when the page is displayed in the browser. Lastly, HTML provides no intrinsic mechanism for including dynamic application data in a web page.

#### 1.1.1 HTML is not a Programming Language
True programming languages, even simple assembly languages, include some basic control instructions like **If** and **Jump**, providing a mechanism to reuse common code. Most assembly languages also provide higher level constructions like **Loop**, **Call** and **Return**, and define reusable code blocks or **Macros**. Higher level procedural languages like C and Fortran, define function and procedure constructs that further facilitate logically organized programming. Finally, object oriented languages like C++ and Java advance these constructs with inheritance and polymorphisms. Unfortunately, HTML lacks even the most basic of these constructs.

#### 1.1.2 HTML Code Lacks Structure and Support for Reusing Code or Variables
Ironically, a web page as viewed from the browser usually has a very well defined and noticeable structure. For instance a typical web page may consist of a header at the top, a footer at the bottom, a navigation column on the left, a promotion column on the right, and a content section in the center. Unfortunately, this well defined structure is completely hidden in HTML code. Typically, the HTML code for a web page consists of one table inside another, inside still another, with the final depth depending on the complexity of the page. This recursive table approach tends to obscure the visually discernible structure of the page.

HTML code structure is further degraded by its lack of reusable components. A typical web page usually contains similar **blocks** like menu items, text sections with captions, input fields with labels, error messages with explanations, links with descriptions, etc. Such blocks merely suggest themselves for being reusable. Unfortunately, repeatable blocks can not be defined in HTML, and such code must be repeated throughout the page.

Similarly, web pages are usually **formatted** or **styled** consistently through out the entire web site. They use the same background, same color scheme, same font faces and sizes. Unfortunately, HTML does not allow for the definition of common format elements.

All of these structural inadequacies combine with very negative consequences for HTML code maintenance.

### 1.1.3 HTML Code Lacks a Dynamic Data Inclusion
The HTML language was basically designed to display predominantly static data in a web browser, yet the vast majority of web applications today need to vary the data they present to a user based upon some business logic. Since it is usually unrealistic to have a static web page available for every possible variation or outcome of a request, most applications have a need to provide dynamic pages, created in real time, with data generated specifically for the user's request. Alas, HTML has no mechanism to include such data.

## 1.2 Compensating Technologies
Attempts have been made to compensate for this essential lack of constructs in HTML. Various technologies like JavaScript, Cascading Stylesheets (CSS), Server Side Includes (SSI), Perl Hypertext Preprocessor (PHP), Active Server Pages (ASP), Servlets and Java Server Pages (JSP), have all been applied widely in the industry, but these solutions are mostly dissatisfying as they are orthogonal to HTML, awkwardly thrust into the HTML code or wrapped around it, rather then intrinsically part of the language. Of all of the compensating technologies, only Extensible Stylesheet Language (XSL) offers a more holistic approach to this problem.

The following subsections examine the more prevalent technologies along with their drawbacks. The final section presents the benefits of XSL for this task.

### 1.2.1 JavaScript
JavaScript is a fully defined programming language that contains all the constructs basic HTML lacks. In addition it includes an object model that completely defines all of the HTML tags on a web page, representing each as an object with functions and properties. Although a very powerful technology, JavaScript is generally used to augment an existing HTML page with additional functionality, rather than to generate a web page. Although techniques do exist to generate HTML using JavaScript, these tend to be awkward, complex and cumbersome, to the extent that the practice is typically limited to small portions of a page. Better technologies exist for this task.

JavaScript is also plagued by a lack of standardization. Code written for one browser may not work in another, making its use for general access problematic or its development an onerous task with multiple versions reserved for particular browser types.

### 1.2.2 Cascading Style Sheets (CSS)
Cascading Style Sheets are a technology that allow for the specification of how an HTML document will be displayed. It is a presentation specification mechanism not a programming language, and as such it augments HTML but does not fully compensate for the features HTML is lacking. In addition, one of the more troubling omissions in this technology is the ability to extend one style from another, making reuse of styles rather clumsy.

### 1.2.3 Servlets, JSP, ASP and PHP
All of these server side technologies compensate for HTML's shortcomings by using other high level languages. Servlets are written in Java and contains numerous statements that output segments of HTML. Java Server Pages are Servlets turned inside out, with HTML code containing embedded Java code. Active Server Pages are a mix of Visual Basic Script and HTML, and similarly PHP is a mixture of Perl and HTML. These higher level languages provide the programming constructs and dynamic data inclusion mechanism absent in HTML, but they tend to even further obscure the connection between the code and its corresponding visual structure. Most code written for these technologies is a complex mix of languages, with the various higher level languages interspersed throughout HTML code. Moreover, because of the ability to execute code on both the server and the client, where a code fragment is executing becomes an additional complicating factor. Reading and maintaining such code can become a challenging exercise in context switching, as the logic shifts from HTML to the higher level language, and from the server to the client.

### 1.2.4 Extensible Stylesheet Language Transformation
Extensible Stylesheet Language Transformation is a technology designed to translate one XML document into another, or more generally, translate an XML document into an HTML document. It consists of an XML based language and an interpreter or engine to perform the transformation. The language is complete with basic programming mechanisms, functions, variables, macros, conditional statements, and loops. Although its syntax looks rather clumsy by itself, it is quite compatible with XML, and provides convenient facilities to access external XML data. Because of the existence of many XML generating and reading technologies available in Java, XSL is ideally suited for an XML emitting servlet design, where the servlets act as the application tier, containing the business logic and outputting data in the form of an XML document, and the XSLT engine acting as an user interface tier, transforming the XML data into an HTML page. Unlike the other server side technologies discussed above, there is no intermixing of technologies. The servlets are written exclusively in Java, the XSL transformation instructions written exclusively in XSL. Like most properly tiered designs this allows for the specialized development, where Java programmers can maintain the application layer, and XSLT programmers can maintain the user interface. Finally, as will be presented in the next sections, XSLT is uniquely suited for logically associating a web page's visual structure with it corresponding code.

## 2. THE DESIGN

As alluded to above, the alternate approach can be based upon Extensible Stylesheet Language Transformation technology. In its simplest form, XSLT relies upon two XML documents, the **data** and the **transformation instructions**. The XSLT engine then takes both these documents as input, and uses the transformation instructions to convert the data to another format, i.e., another XML document, an HTML document, or any format that can be specified in the transformation instructions. The alternative approach presented here goes further, however, defining a common XSL library to be applied to two XML documents, one of which contains web page templates and the other contains the dynamic data. It is this approach that more adequately compensates for all the features HTML lacks.

Specifically this approach will support the following features:
- Concise, readable web page definitions.
- Componentized, adjustable web page macros that support variable parameters.
- Global style and format definitions.
- Dynamic data inclusion.

The following sections will define and detail this approach.

### 2.1 Web Page Structure

The strongest feature of XSLT-based web development is its ability to define a web page structure in XML. Such a structure is presented in the code below.

```
<Page name="search.html">
 <Header>Search</ Header >
 <Menu>
  <MenuItem link="index.html">Home</MenuItem>
  <MenuItem link="search.html">Search</MenuItem>
  <MenuItem link="help.html">Help</MenuItem>
 </Menu>
 <Content>
 <Title>Customer Search</Title>
 <Text>All fields are optional.</Text>
 <Form action="results.html">
    <InputField label="First Name:"  prop="Customer.FirstName"/>
    <InputField label="Last Name:"  prop="Customer.LastName"/>
    <InputField label="State:"  prop="Customer.Address.State"/>
    <InputField label="Zip:"  prop="Customer.Address.Zip"/>
    <Submit label="Find"/>
 </Form>
 </Content>
 <Footer>OOML © 2008</Footer>
</Page>
```

In a few lines it describes all the essential elements of this particular web page, a **menu** containing three items that reference other web pages, a **title** on the main section, some explanatory **text**, and a **form** with the four **input fields** along with their associated **labels** and CGI parameter **names**. Note this web page definition is devoid of any presentation information, focusing only on structure. This simple paradigm enables the easy creation of new web pages and facilitates high-level changes to existing web pages. For instance, adding another menu item to that page or changing the footer is a rather intuitive task. Similarly, adding or modifying an input field is quite easy.

### 2.2 The OOML Language

As is plainly seen, there is no HTML in the above example. The web page structure is completely defined in a language other than HTML. The definition consists entirely of XML tags. Each of these tags corresponds to an XSL template, a set of XSL instructions, where the target HTML presentation is defined. To render a particular web page, an XSL transformation is performed on the web page structure definition using these predefined instructions. Although one could argue that the complex and awkwardness of HTML is simply being shifted to XSL templates, there are significant benefits in this indirection. Once a core set of XSL instructions is defined, they can be offered as a library and reused readily from application to application, so that, in a very real sense, this structure definition is the basis for an entirely new markup language. What remains is the introduction of the other promised features, macros, control constructs, dynamic data access, etc. As will be shown in the next sections, these inclusions merely involve adding XSL instructions to the library.

### 2.3 Macros

As discussed, **macros** are reusable code blocks that allow commonly used code segments to be defined and referenced throughout a larger application. This is a powerful mechanism as it allows for the definition of larger, aggregated functional units which gives rise to conceptually simpler code. There are also added advantages to larger, more sophisticated applications, where macros definitions can be grouped logically and segregated into certain designated sections of code or stored in separate files. Such organizations facilitate code maintenance.

Macros can be added to this emerging markup language simply by adding additional translation instructions to the XSL library.

The syntax for macro definitions is illustrated below.

```
<Macro name="CustomerSearchForm">
 <Title>Customer Search</Title>
 <Form action="results.html">
    <InputField label="First Name:"  prop="Customer.FirstName"/>
    <InputField label="Last Name:"  prop="Customer.LastName"/>
    <InputField label="State:"  prop="Customer.Address.State"/>
    <InputField label="Zip:"  prop="Customer.Address.Zip"/>
    <Submit label="Find"/>
 </Form>
</Macro>
```

In this example, the macro defines a customer search form. Now, instead of explicitly specifying the form, the web structure definition may call this **CustomerSearchForm** macro, and, if needed, other similarly defined macros as the following code fragment illustrates.

```
<Content>
  <Call macro="CustomerSearchForm"/>
  <Call macro="CustomerSearchResults"/>
  <Call macro="CustomerSearchHelp"/>
</Content>
```

### 2.4 Macro Parameters

The ability to pass parameters to a macro definition greatly enhances its capabilities. Parameters allow the macro to be adjusted or customized for a particular use. The emerging OOML language presented here supports the passing of a single value, a set of values, or even XML code as a parameter to a macro call. Such macro call parameters are to be placed within the **<Call>** and **</Call>** tags, and should be in XML format. For example, a single value parameter called X should be passed as the following:

```
<X>100</X>
```

If multiple values are to be passed, for instance City, State, and Zip, they can be optionally aggregated into a larger structure and passed to a macro, as the following code illustrates:

```
<Call macro="Content">
  <Address>
    <City>Middletown</City>
    <State>NJ</State>
    <Zip>07748</Zip>
  </Address>
</Call>
```

The actual parameters will be available in the macro definition by means of the <Param> construction.

```
<Param select="Address.State"/>
```

Note the name convention employed is quite similar to that of XML Path Language (XPath) expressions. The XML elements are referred to by a hierarchical series of dot-separated, names. (As will be shown later, this same convention will also be used to access dynamic data and to parse CGI parameters.)

As mentioned above, XML code can be passed as a macro parameter. This is quite different from passing data, as the code is to be *interpreted* in the macro. The goal here is to make macros as powerful as XSL templates, and, generally, the definition of macros is preferred over the creation of new custom XSL templates.

In the previously defined web page structure, the **Content** specification can be replaced with a macro. The macro call will look like the following.

```
<Call macro="Content">
  <Call macro="CustomerSearchForm"/>
  <Call macro="CustomerSearchResults"/>
  <Call macro="CustomerSearchHelp"/>
</Call>
```

In this code segment, the XML block consisting of the three internal macro calls will be passed to the macro Content as a parameter. The called macro can proceed with the interpretation of the passed code by means of the **<Continue/>** statement.

If in the above example the Content macro is defined as the following.

```
<Macro name="Content">
  <Box>
    <Title>Help Desk Application</Title>
    <Continue/>
  </Box >
</Macro>
```

The macro will draw a box, display the title, and proceed to interpret the XML code passed from with in the macro call. In this case, it will just call the three specified macros one by one.

## 2.5 Dynamic Data

Conventionally, a web application, when invoked, accepts parameters, performs business logic, calculates or generates some dynamic data, and composes a web page using this data. In this new approach, it will be necessary for the application to issue its dynamic data in the form of an XML document. In this way, the same XSL transformation used to render an HTML document can be employed to populate it with dynamic data.

Like the macro data parameters describe above, the dynamic data may consist of plain string parameters like "<X>100</X>," or of aggregated data. For example, Customer Data may consist of Personal Information and an Address. The Personal Information may further consist of First Name and Last Name. The Address, as in the example above, may consist of Street, City, State and Zip. The corresponding XML document might look like the following:

```
<Node>
  <Data>
    <Customer>
      <Person>
        <FirstName>Joe</FirstName>
        <LastName>Boxer</LastName>
      </Person>
      <Address>
        <City>Middletown</City>
        <State>NJ</State>
        <Zip>07748</Zip>
      </Address>
    </Customer>
  </Data>
</Node>
```

To access the dynamic data the same convention of dot-separated names used in referring to macro parameters and, as will be shown later, parsing CGI parameters, will be employed. The library provides the following construction to insert a dynamic data element:

```
<Value select="Customer.Address.State"/>
```

This refers to the following data element in the passed XML document:

```
<Customer>
  <Address>
    <State>NJ</State>
  </Address>
</Customer>
```

Below is another example of referencing to dynamic data in an input field.

```
<InputField label="State:"  prop="Customer.Address.State"/>
```

## 2.6 Control Constructions

The library provides constructions to conditionally include or omit OOML code, based upon equal or not equal statements. These conditional statements are dependent on the dynamic data values, which are referenced using the same XPath like convention described above. The following code fragment illustrates a conditional case based upon the State information in the customer's address:

```
<Ifeq select="Customer.Address.State"  value="NJ">
  ...
</Ifeq>
```

This statement will access the data element specified by the **select** attribute and check whether or not it is equal to the string given in the **value** attribute. The body of this construction will be processed or skipped accordingly.

The **<Ifne>** statement with the similar syntax implements the opposite condition.

## 2.7 Dynamic Data Blocks

Dynamic data has transaction scope, in that it exists and is unchangeable in the context of the XSLT transform operation. Thus, for the purposes of the OOML coding they are very similar to global variables, accessible anywhere in the code.

The library also provides a variable localization feature. The **<Block>** statement can be used to specify a base reference point for all dynamic data specification within the scope of the tag. For instance, if a section of code deals mainly with customer data, the following **<Block>** could be employed:

```
<Block base="Customer">
    ...<Value select="Address.State"/>...
</Block >
```

Inside the Block structure the **Customer** portion of the XPath reference is omitted on all dynamic data references.

This construct can be especially useful in adjusting macro calls. In cases where the dynamic data tree has nodes with similar sub-trees, the **<Block>** statement can be used to set different base points, allowing the same included constructions and macros to produce different results.

```
<Block base="Customer">
    ...<Value select="Address.State"/>...
</Block >
    …
<Block base="Vendor">
    ...<Value select="Address.State"/>…
</Block >
```

## 2.8  Loops
The same idea allows the definition of loops based on the dynamic data. The **<ForEach>** tag enables the repetition of OOML code based upon multiple instances of elements within the dynamic data. In the following example, the code contained in the **<ForEach>** element will be repeated for each instance of a Customer element in the dynamic data.

```
<ForEach select="Customer">
    ...<Value select="Address.State"/>...
</ForEach >
```

## 3.  THE PRESENTATION
The previous section introduced the programming construction lacking in HTML. They form the highest logical layer in OOML. In this section the next layer will be introduced, a **Presentation** layer. Using the same XSLT technique, the language will be expanded with more constructs, forming the basis of an entirely new higher level language that has the potential to supersede HTML for web page development.

This presentation layer can be logically divided into two sub layers, one specifying where web page elements should be placed, referred to as the **Layout** layer, and another specifying how elements should look, referred to as the **Style** layer.

## 3.1  Layout
In HTML, one of the most common mechanisms used to define layout involves the TABLE element. The TABLE paradigm is a fairly complex HTML construction, with embedded Table Row (<TR>) elements, which, in turn, contain Table Data (<TD>) elements that actually hold the elements to be displayed. Although this TABLE paradigm is a very powerful construction, with capabilities to define virtually all possible layouts, it is also one of the most difficult to maintain, especially if one table is inserted into another, as is commonly done. Because of this table nesting, it is often very difficult to fix even the simplest of errors or make seemingly trivial alterations.

Another common HTML layout mechanism employs the DIV element. The DIV element enables the division of a web page for the purposes of defining common generic properties within that division. DIVs are easier to use than TABLEs, as they do not have embedded rows and columns. They seem to be well suited for defining fixed width web pages using absolute positioning, but they can become very complicated when developing web sites with floating or relatively positioned elements.

In this initial instantiation of OOML, the power of HTML TABLES and the flexibility of DIVs, did not outweigh the need to simplify development while maximizing functionality. Instead the approach defines a few simple constructions that capture the most commonly used layouts. More can certainly be added as this technology matures.

Ironically, the low-level implementation of the offered layout constructions may employ both TABLEs and DIVs, but this use will be completely obscured in the OOML code. This indirection also allows the low level implementation to change when needed, without impacting the OOML coded web pages.

Currently, there are three layout constructions, **ListV**, **ListH**, and **LeftRight**. These are simple XML constructions, consisting only of opening and closing tag brackets, but they manage a number of included constructions. The ListV construction places its sub-elements vertically. The ListH construction places its sub-elements horizontally. The LeftRight construction expects exactly two sub-elements. It places them to the left and to the right, in a slightly different manner than a construction using the ListH element with two sub-elements, which places the second element nearer to the middle.

In the example below, the left and right plain text elements are explicitly specified as such, i.e., the text "Left" and the text "Right".

```
<LeftRight>
    <Text>Left</Text>
    <Text>Right</Text>
</LeftRight >
```

Although it is possible to nest layout constructions to further specify more complicated layouts, the constructs tend to be less complex that their TABLE counterparts, as the specifications are more accurately and simply defined, and retain an intuitive connection to the visual structure of the web page.

## 3.2  Style
Whereas the layout constructions define where an element should be displayed, the style constructions define how the element should look. In HTML it is possible to define the style of each element, including its color, font, size, etc, but these specifications are too fine grained. For the sake of simplicity, it is desirable to deal in more large grained constructs, plain text, title, or super title, each of which has a unique and distinctive appearance that is consistently applied thought the entire web site. To this end, OOML typically defines style constructs like the **Text**, **Title**, and **SuperTitle** elements. The following code fragment illustrates their use.

```
<SuperTitle>
    Section 2.  THE DESIGN
</SuperTitle>
<Title>
    Section 2.1.  Web Page Structure
</Title>
<Text>
    …
</Text>
```

Just as in the layout constructions, the actual HTML implementation is not defined in the OOML code, but is relegated to the XSLT template defined for that construction. Again, this level of indirection allows the actual style to change without impacting the OOML code defining the web page.

It is also common to visually group elements on a web page by enclosing them in a box. Such a construction also exists in OOML as the **Box**. A sibling construct, the **Panel**, is a box containing a title. A Panel code fragment is presented below.

```
<Panel title="Error Message">
    <Text>…</Text>
</Panel>
```

The **NoBox** construction is provided for the sake of completeness. This is a box with an invisible frame, where the invisible frame occupies exactly the same space as a visible one would. This construct can be useful in rare occasions.

### 3.3 Atoms

Atoms refer to higher level stylistic elements, usually in the form of compound or complex user interface controls, like input fields, hypertext links, menu items, etc. Structurally they are no different from other stylistic constructions as they are all defined by XSL templates.

Currently, the following atoms are supported in OOML:
- AtomInputField (see above,)
- AtomPassword,
- AtomHidden,
- AtomTextArea,
- AtomSelectState,
- AtomYesNo,
- DropDownList (see below,)
- RadioList,
- AtomCheckBox,
- AtomSubmit

The code segment below illustrates a typical use of Atoms in OOML.

```
<DropDownList label="Phone Type:"
  prop="Customer.Phone.Type">
    <Option id="" label=""/>
    <Option id="H" label="Home"/>
    <Option id="W" label="Work"/>
</DropDownList>
```

The code above will result in the creation of an HTML SELECT element, with three embedded OPTIONS, an empty option, Home and Work, and a name attribute of Customer.Phone.Type. With minimal decoration the result will look like the following.

```
<EM>Phone Type:</EM><BR>
<select name="Customer.Phone.Type">
  <option value=""></option>
  <option value="H"
      default="yes">Home</option>
  <option value="W">Work</option>
</select>
```

Notice that the default value will be set to a particular option depending on the dynamic data element specified by the "prop" attribute, namely by **Customer.Phone.Type** in this case.

```
<Customer>
  <Phone>
    <Type>H</Type>
  </Phone>
</Customer>
```

The same is also true for RadioList and AtomSelectState atoms.

Of course, if needed, other atoms (or macros) can be defined.

### 3.4 Global Style Variables

All style constructions are parameterized via global XSL variables, which define such things as the color scheme, font faces, and graphics for the whole web site. These global variables are not hard coded in XSL instructions. It is more convenient to define them in an external XML file, and then use that file to initialize the global XSL variables. Moreover, the global style values can be conditionally set, allowing a general look and feel to be altered depending upon some characteristic of the site such as the domain name.

## 4. THE IMPLEMENTATION

Briefly, the proposed approach employs XSL transformation to render HTML pages based on their concise definitions in OOML, a markup language supporting basic text processing and dynamic data access features. The OOML language is implemented as a library of XSL templates. Together with an XSLT engine, the provided XSL library forms a type of OOML translator applicable to generate HTML pages.

### 4.1 Client Side or Server Side Transformation

In addition to a number of XSLT engines available for use in the application tier, most modern browsers are equipped with embedded XSLT engines. This gives rise to two fundamental options, perform the XSL transformations on the server or perform them on the client.

Although it is tempting for application developers to off-load this task from what might be a busy application server to the client's desktop, experience has shown that server side transforms are the more prudent choice. Client side transformations depend too heavily on the browser type and version, and often lead to inconsistent results. Server side transformations, with the XSLT engine specified by the application, are far more reliable.

### 4.2 Revealed Power of XSL

XSL is rarely used for web development not only because of its unusual syntax, but also due to a lack of functional examples. Thus the Xalan XSL engine package includes an example of performing XSL transformation in a servlet. The Xalan example requires two static input files, the XML data and the transformation instructions designed to act on that data. As a result, the example servlet produces an HTML page. In terms of functionality, this is equivalent to serving a static page.

This example may be considerably improved by simply allowing the XSL code to access dynamic data. To this end, the dynamic data should be converted to XML format first. Then the same servlet could perform the XSL transformation on the dynamic XML (see below) and the static transformation instructions.

```
<Screen>
  <Config>
    <IP>127.0.0.1</IP>
    <Page>index.html</Page>
  </Config>
  <Input>
    <X>100</X>
  </Input>
</Screen>
```

The result will be a dynamic HTML page. The content of this page may depend on such dynamic data as the domain name, the page name, the time, the client IP, etc. Moreover, the modified example servlet may parse input CGI parameters and add them to the dynamic XML in the same manner so that they may be accessible in the XSL code as well.

These minor modifications turn XSL into a web scripting language, capable of implementing certain business logic. The semi-dynamic (or quasi-static) web pages can display different content, depending on the required page. It can also make simple calculations over the input parameters and display the results. It can search an external XML file against certain criteria based on the input parameters or calculations. Theoretically it can even write something to an external XML file on the server.

Though this technique demonstrates the little-known power of XSL, and will be good for rendering web pages with limited functionality like changing appearance, showing different static information, or hiding some parts, depending on the user input or HTTP parameters, we do not recommend XSL as a programming language for complicated web applications. Extensive business logic should be implemented in a more convenient language, like Java.

### 4.3 Structure Separation

As assumed above, all XSLT applications involve two logical inputs to the XSLT engine, *data* and *transformation instructions*. The data is the application data intended for display. The transformation instructions are the instructions to translate that data into a web page. Typically, the data does not specify the web page structure, but rather this structure is hard coded into the translation instructions. This is powerful, but rather rigid, as it requires the transformation instructions be changed each time there is an alteration to the page structure. As shown above, one of the most powerful aspects of OOML is its ability to define web page structure in a form of the XML data. However, it would be very cumbersome for the application tier of a system to produce this web page structure data as part of its output. Indeed, the principles of three-tiered architecture prohibit such intermingling of application and presentation logic. This problem is avoided in OOML by isolating web page structure definitions to separate, static files in the web application. These files are then referenced by the transformation instructions. This is a powerful construct, as it allows an application's entire web page schema to be changed simply by changing the page definitions referenced. The application tier would not be impacted. It also greatly aids code development, as it cleanly divides application logic from presentation logic, with application developers focusing on producing the output data, and presentation developers focusing on defining web page structures.

To put it succinctly, where typical XSLT applications have two logical inputs to the XSLT engine, *data* and *transformation instructions*, OOML applications have three, *data*, *structure* and *transformation instructions*.

### 4.4 Parse the Structure – Reference the Data

In the OOML paradigm, there are two XML data documents to parse, the application data and the web page structure. This creates a logical problem, since the XSTL engine expects only one. OOML deals with this issue by opting to parse the web page structure definition and reference the application data. This logical pivot from application data to page structure is accomplished by the following two XSL statements.

The first statement switches parsing from the input application data to an external XML source, in this case the web page structure definition.

```
<xsl:template match="/">
  <xsl:apply-templates
    select="document('pages.xml')/*[...]"/>
</xsl:template>
```

Where the **pages.xml** reference is a static file holding the set of web page definitions.

The second statement preserves input application data by creating a global reference to it. This global reference can then be used through out the XSL templates to reference this dynamic data.

```
<xsl:variable name="node" select="/*"/>
```

Where the XSL variable **$node** now points to the root of the dynamic data document.

### 4.5 Web Page Generation

The OOML implementation is comprised of an XSLT servlet and the XSL library mentioned above. Given a data document and web page specifications, a web page is generated. This is accomplished by invoking the appropriate macros, observing conditional constructions, interpreting style elements, and accessing dynamic data, all of which is supported by the provided XSL library.

Taking advantage of the XSL include statement, the XSL library actually consists of one main XSL file and a number of auxiliary XSL files, with auxiliary files containing logically grouped transformation instructions. The main XSL file then references these auxiliary files, defines a number of global variables, and defines a root template where the transformation begins. As detailed in the previous section, this root template is also responsible for pivoting the XML parsing to the web page definitions and establishing the global pointer to the application data.

In current instantiation of OOML, the page structure definitions are assumed to be put into the **pages.xml** file. Alternatively, the presentation programmer can alter the main XSL file to refer to a different page structure definition file.

The XSLT servlet parses CGI and HTTP parameters and the page name part of the requested URL, e.g. "index.html." The page name is used to search for the particular page definition with the name attribute equal to "index.html" in the **pages.xml** file.

```
<Pages>
 <Page name="index.html">
   <!-- OOML code here -->
   ...
 </Page>
 <Page name="search.html">
   ...
 </Page>
</Pages>
```

### 4.6 A Functional Web Application

The goal of OOML is to facilitate web page development, by utilizing well known programmatic techniques and approaches like data variables, control constructions, macro generation, code reuse, etc. Sometimes web pages written in OOML can be used for informational web sites, but more often they are to form a GUI to a web application. In the latter case OOML becomes a part of a larger application framework. It is conventional to distinguish the following three tiers in a web application: front-end, application server, and the database. The OOML forms the front tier.

It is important for the front tier to be well separated from the rest of the application. When the front end is written in OOML, it contains all the presentation information and fully controls how to display and decorate the output data, and what it requires from the application tier is only the values of the data . The application tier is responsible for performing the required business logic and emitting the data to be displayed in the XML format. The front tier is fully and solely responsible for displaying it.

If the middle tier is implemented in Java, the servlet should parse CGI parameters, perform the business logic, render the result to XML, and perform the XSL transformation to translate OOML. All this should not necessarily be done in one servlet. Namely, all interface related actions may be done in a general servlet, while particular business logic may be moved to a separate class.

This approach is further developed in so-called Summer MVC tool, and will be covered in full details in the next article.

## 5. XSL TRICKS
This section is provided to illustrate some powerful, though not intuitive, features of XSL, some of which were utilized in OOML.

### 5.1 The Dynamic XPath
The XSL language allows access to data elements from an XML source by XPath expressions. Static XPath expressions are hard coded in XSL templates.

```
<xsl:value-of select="//Customer/Address/State"/>
```

Often this is not enough. There may be a need to refer to data by a Dynamic XPath expression, which itself is taken from some XML element or attribute. Dynamic XPath expressions are not directly supported by XSL, but can be implemented by means of a recursive XSL template. The auxiliary template parses the XPath value given as a string parameter for the first element and the rest. Then it selects the appropriate data sub-tree from the source XML data and recursively calls itself to go down the selected sub-tree along with the rest of the desired XPath.

### 5.2 Switch to External XML Code
The transformation starts with the main XSL file. The root XSL template fetches the name of the required page, "search.html," from the Config section, and passes XSL control to the external file (pages.xml) containing the structure definitions of the web pages, namely to the part (Page) of that file with the corresponding name attribute.

```
<xsl:variable name="page">
   <xsl:value-of select="/Screen/Config/Page"/>
</xsl:variable>

<xsl:template match="/">
<xsl:apply-templates
   select="document('pages.xml')//Page[@name=$page]/*"/>
</xsl:template>
```

As stated, the **pages.xml** file contains the structure definitions for the web pages, and it was simply convenient to keep all such definitions in one file.

### 5.3 Accessing External XML Data
The same approach can be applied for obtaining data defined in an external XML source.

The following demonstrates how to select a set of global variables depending on a domain name configuration value.

```
<xsl:variable name="lnf">
    ../view/lnf.xml
</xsl:variable>
<xsl:variable name="domain">
    <xsl:value-of select="//Config/Domain"/>
</xsl:variable>

<xsl:variable name="text.color">
   <xsl:value-of
    select="document($lnf)//Scheme[@name=domain]
    /text/@color"/>
</xsl:variable>
```

## 6. CONCLUSIONS
This paper examined the various drawbacks inherent in the Hypertext Mark up Language as it is currently employed in Web development. It then proposed an alternative approach using server-side, Extensible Stylesheet Language Transforms (XSLT). In this alternative an entirely new, higher level, markup language was created, which contained the constructs missing in HTML and enabled web page specifications that are far more intuitive then those in HTML.

Although the paper focused on OOML as a viable alternative to HTML, the mechanism is actually a part of a larger Java application framework that provides the accompanying input and output marshalling and server side XSLT transformation alluded to in the implementation section. The authors plan to present this application framework in future publications.

## 7. ACKNOWLEDGMENTS
We would like first to express our sincere appreciation to Kenneth Gaylin for the excellent job in correcting and proof-reading the manuscript. We also express our special thanks to Mr. Melvin McArthur for releasing this paper for publication.